\documentclass[prd,aps,showpacs,tightenlines,superscriptaddress,amsmath,amssymb,amsfonts,twocolumn]{revtex4}
\usepackage{amssymb,amsmath,amsthm,amsbsy,epsfig,color,graphicx,times}
\usepackage[ansinew]{inputenc}
\usepackage[english]{babel}
\usepackage{color}
\usepackage{braket}
\usepackage{tabularx}
\usepackage{array}

\def\be{\begin{equation}}
\def\ee{\end{equation}}

\def\bea{\begin{eqnarray}}
\def\eea{\end{eqnarray}}

\begin{document}

\title{ Neutron Interferometry and axion like particles}

\author{Antonio Capolupo}
\affiliation{Dipartimento di Fisica ``E.R. Caianiello'' Universit\`{a} di Salerno, and INFN --- Gruppo Collegato di Salerno, Via Giovanni Paolo II, 132, 84084 Fisciano (SA), Italy}

\author{Salvatore Marco Giampaolo}
\affiliation{Institut Ru\dj er Bo\v{s}kovi\'c, Bijeni\v{c}ka cesta 54, 10000 Zagreb, Croatia}

\author{Aniello Quaranta}
\affiliation{Dipartimento di Fisica ``E.R. Caianiello'' Universit\`{a} di Salerno, and INFN --- Gruppo Collegato di Salerno, Via Giovanni Paolo II, 132, 84084 Fisciano (SA), Italy}

\begin{abstract}
We propose a method to reveal axions and axion--like particles based on interferometric measurement of neutron beams. We consider an interferometer in which the neutron beam is split in two sub--beams propagating in regions with differently oriented magnetic fields. The beam paths and the strength of the magnetic fields are set in such a way that all the contributions to the phase difference but the one due to axion--induced interactions are removed. The resulting phase difference is directly related to the presence of axions.
Our results show that such a phase is in principle detectable with neutron interferometry, possibly proving the existence of axions and axion--like particles.
\end{abstract}

\maketitle

The need for physics beyond the standard model of particles \cite{Ellis2009} is evident in several phenomena, from neutrino mixing \cite{Mixing1,Mixing2,Mixing3,Mixing4} to the dark sector of the universe \cite{dark1,dark2}. A hint of new physics also comes from strong interactions, for which the standard model allows an arbitrary CP symmetry violation, through the $\theta$ term, that is not observed experimentally. This tension, known as the strong CP problem, can be elegantly reconciled by promoting the $\theta$ parameter to a dynamical field, as originally proposed by Peccei and Quinn \cite{Quinn1977,Peccei1977}. The Peccei--Quinn mechanism gives rise to pseudo--scalar particles known as axions \cite{Quinn1977,Peccei1977,Weinberg1978,Wilczek1978,Raffelt2007} which hold a number of interesting properties. In particular, they soon gained attention as a possible dark matter component \cite{Marsch2016} in virtue of their tiny interactions with ordinary particles. Inspired by the Peccei--Quinn model, a considerable variety of axion--like particles (ALPs) models has been proposed, including the DSFZ~\cite{DFSZ1, DFSZ2, PDG2020} and the KSVZ axions~\cite{KSVZ1, KSVZ2, PDG2020}, with masses ranging everywhere from the ultralight \cite{Kim2016,Demartino2017} $m \simeq 10^{-22} \mathrm{eV}$ to the heavy axions \cite{Rubakov1997} $m \simeq 1 \ \mathrm{TeV}$. This proliferation of models is justified, at least partially, by the primary role that ALPs would play in cosmology. Indeed, ALPs represent to date one of the most convincing particle explanation for dark matter and are actively searched for in several experiments \cite{PVLAS,OSQAR,OSQAR2,ALPS,CAST,ADMX,QUAX,Capolupo2015,Capolupo2019}. Nonetheless, up to now no evidence for ALPs has been found.

Besides the interaction with the electromagnetic field, on which the vast majority of the ongoing experiments is based, ALPs are expected to experience an interaction with fermionic fields. The latter yields an effective fermion--fermion interaction \cite{Moody1984,Daido2017} in which the ALP plays the role of a mediating boson. For this reason, the detection of ALPs is naturally tied not only to electrodynamics but also to the possibility of experimentally probing interactions between fermions \cite{Bezerra2014,Klimchitskaya2015,Klimchitskaya2017,Capolupo2020}. On the other hand, in the context of neutron physics, interferometry has proven to be a formidable tool in investigating particle interactions and their quantum properties. Neutron interferometry is indeed a subject of study in its own right \cite{Rauch2015} and has granted us the chance to verify many theoretical predictions. These include the Sagnac effect \cite{Werner1979}, the geometric phase \cite{Allman1997,Wagh1990,Wagh1998}, wave--particle duality and the change of sign in the spin $1/2$ wave--function after a $2 \pi$ rotation \cite{Rauch2015}.

The purpose of this letter is take advantage of the versatility of neutron interferometry in order to provide an alternative approach to the detection of ALPs. We show that the axion--induced interactions between neutrons sum up to give an additional magnetic term, whose presence affects the phase of the neutrons and produces an extra phase difference in interferometric experiments.
We consider an idealized interferometric setup in which a collimated neutron beam is split into two sub--beams that are later let interfere with each other. Each of the sub--beams is subject to an external magnetic field of equal strength but different direction. In particular,
if one of the two magnetic fields is set in the direction of propagation of the relative sub-beam and the other is orthogonal to the propagation, the neutrons will gain a path-dependent phase factor that can be easily detected by an interferometer.
The experimental parameters are fixed in such a way that the interference pattern is only due to the magnetic and axion--induced interactions. We show how to isolate the axion contribution by   appropriately choosing  the neutron path and we set an arm length such that the contribution to the phase difference given by the  dipole-dipole interaction is an integer multiple of $2 \pi$.
In this way, we obtain  a phase difference depending only on the axion--mediated interaction. We discuss how the presence of ALPs can be tested with a neutron interferometer in a significant portion of parameter space, thus proposing  a new tool for the search for one of the most elusive particles in the universe.

The Lagrangian describing the coupling of the neutron field $\psi$  with the axion $\phi$, is given by
$
  \mathcal{L}_{INT} = -  i g_{p} \phi \bar{\psi} \gamma_{5} \psi
$
where $g_{p}$ is the (dimensionless) effective axion--neutron coupling, which depends on  the axion (or ALP) model \cite{Moody1984,Daido2017}. The couplings are generally small $g_{p} \ll 1$, allowing a perturbative treatment of the interaction.  We shall assume that the neutron velocities are non-relativistic and analyze this interaction potential within the context of ordinary quantum mechanics. In the non--relativistic limit $\mathcal{L}_{INT}$ yields a two--body potential for the neutrons \cite{Moody1984,Daido2017}. This axion--induced interaction is not the only one in play. However, the gravitational interaction is easily seen to be irrelevant, due to the smallness of the masses involved. In addition, we shall always assume a relative distance $r > 10^{-12} m$ among the neutrons, so that the short--range nuclear interactions can also be ignored. With this assumption the only other relevant interaction is the magnetic one between the neutron dipoles.
The two--neutron interaction Hamiltonian, comprising the magnetic and the axion--mediated interaction can thus be written as
\begin{eqnarray}\label{Total_Hamiltonian3}
\nonumber H_{ij} &=& -\frac{\mathcal{A}}{r^3_{ij}} \bigg[ \left(3-\mathcal{B}e^{-mr_{ij}} K^{(a)}(r_{ij})\right) \sigma^{r_{ij}}_i \sigma^{r_{ij}}_j \\
&-& \left(1-\mathcal{B}e^{-mr_{ij}}K^{(b)}(r_{ij})\right) \pmb{\sigma}_i \cdot \pmb{\sigma}_j \bigg] \ .
\end{eqnarray}
In eq. \eqref{Total_Hamiltonian3} the two contributions are signaled by the parameters  $\mathcal{A}=\frac{g^2 \alpha}{16 M^2}$, denoting the strength of the magnetic interaction ($g$ is the neutron g--factor, $M$ is the neutron mass and $\alpha$ is the fine-structure constant) and $\mathcal{B}=\frac{4 g_p^2 }{g^2 q_e^2}=\frac{ g_p^2 }{ \pi \alpha g^2}$, representing
the relative weight of the axion interaction, which vanishes in absence of the ALP (whose mass is denoted $m$). The dimensionless (in natural units $c=1=\hbar$) functions $K(r)$ are defined as $K^{(a)}(r) = m^2 r^2 + 3m r + 3$, and $K^{(b)}(r) = mr + 1$. The vector $\pmb{r_{ij}} = \pmb{r_i} - \pmb{r_j}$ denotes the relative position of the neutrons $i$ and $j$, $r_{ij}=|\pmb{r_{ij}}|$ is their relative distance and the operators $\sigma^{r_{ij}}_l = \pmb{\sigma}_l \cdot \pmb{\hat{r}_{ij}}$ are defined by the projection of the pauli operators $\pmb{\sigma_l}$ of neutron $l$ on $\pmb{r_{ij}}$. The notation $\sigma^{r_{ij}}_i$ is used to remark that while these operators act only upon the space of the $i$-th particle, their form depends on the specific pair $i,j$ considered.  Eq.\eqref{Total_Hamiltonian3} can be recast in a more compact form by defining the functions $ C (r) = \frac{\mathcal{A}}{r^3}\left(1 - \mathcal{B} e^{-mr} K^{(b)}(r) \right)$, $D(r) = \frac{\mathcal{A}}{r^3}\left(3 - \mathcal{B} e^{-mr} K^{(a)}(r) \right)$ and the symmetric matrix $ K^{uv} (\pmb{r})= C (r) \delta^{uv} - D (r) R^{u} (\pmb{r}) R^{v} (\pmb{r})$ for $u,v = x,y,z$. The symbol $R^{u}(\pmb{r}) = \pmb{\hat{r}} \cdot \pmb{\hat{u}}$ denotes the projection of the vector $\pmb{r}$ on the $u$ axis. Thus $H_{ij} = \sum_{u,v} K^{uv}(\pmb{r}_{ij}) \sigma_{i}^{u} \sigma_{j}^{v}$. The two--neutron interaction can easily be generalized to an arbitrary number of neutrons. The total interaction Hamiltonian is simply the sum over all pairs $i,j$
$
  H = \frac{1}{2}\sum_{i,j} H_{ij}
$
, where the factor $\frac{1}{2}$ accounts for double counting.

The evolution of the $n$--neutron state with a generic $n$, interacting via $H$ is in principle a complicated many--body problem, which hardly admits an analytical treatment if no further assumptions are made. In the present context, however, we are not interested in correlations and collective effects, and are concerned only with the evolution of the single neutron state. For this reason it is convenient to encode the interaction with all the other nucleons in an effective one--particle potential, by resorting to a mean field approach. Given a neutron at position $\pmb{r}_i$ and pauli operator $\pmb{\sigma}_i$, the instantaneous interaction hamiltonian due to the other neutrons is
$
  H_i = \sum_{u,v}  \sum_{j \neq i} K^{uv} (\pmb{r_{ij}}) \sigma_{j}^{u} \sigma^{v}_{i} \ ,
$
where the sum runs over all the other neutrons $j$. To obtain an effective local potential for the single neutron, this equation is replaced with its expectation value on the state of the other nucleons. In doing so, the single particle Hamiltonian $H$ acquires a very neat interpretation in terms of an effective magnetic field due to the other neutrons. Indeed, setting
\begin{equation}
 \mu \pmb{B_i} (\pmb{r_i}) =-\sum_u   \sum_{j} K^{uv} (\pmb{r_{ij}}) \langle \sigma_{j}^{u} \rangle \,,
\end{equation}
we recover the usual term of a spin interacting with a magnetic field
$
  H_{i} = -\mu\left(\pmb{B_i} (\pmb{r_i})\right) \cdot \pmb{\sigma} \ .
$
Once the effective magnetic field is computed for a particular spatial--spin configuration, one simply plugs the one--particle operator in the Schroedinger equation in order to study the evolution of the single neutron state.

For our purposes, we shall consider neutron beams with specific requirements. First of all, we deal with collimated neutron beams. For cold neutrons, collimation and a small beam width of the order of $10 \ \mathrm{\mu m}$ can be obtained by any of the means described in the reference \cite{Ott2015}. Neglecting the angular spread, we can generally consider that the beam is distributed with cylindrical symmetry around the beam axis $\pmb{\hat{y}}$, and, assuming that it is sufficiently thin, we model the beam as a monodimensional system.
We will also assume that only neutrons around a given value of the kinetic energy $K$ are selected, for instance by use of a monochromator. The beam intensity is expected to decay as the neutron beam propagates. However, as a first instance, we neglect the losses due to the propagation, and consider the intensity constant.

Given these preliminaries, we now discuss an idealized setup aimed at revealing the axion--mediated interaction among neutrons. A beam of cold or ultra--cold neutrons, whose source \textbf{SRC} might be an appropriate reactor, is conveyed to an external apparatus \textbf{EXT} which has the purpose of rendering the beam as monochromatic as possible (e.g. using monochromators) and also serves as a collimator, making the beam as thin and linear as possible. The beam then goes through a beam splitter \textbf{BS} which splits the beam into two sub--beams $I$ and $II$. Then each of the sub--beams enters a Stern--Gerlach like apparatus that selects two different spin polarizations $\pmb{P}_{I}$ and $\pmb{P}_{II}$ (the upper case indices $J = I,II$ and roman numerals are used to denote the specific sub--beam). The average polarizations are maintained by two constant and uniform magnetic fields $\pmb{B_{I}^{0}}$ and $\pmb{B_{II}^{0}}$ surrounding the sub--beams $I$ and $II$ of the same strength but with distinct direction, i.e. $\pmb{B^0_J}=B_0 \pmb{P_J}$.

\begin{figure}[t]
\centering
\includegraphics[width=\linewidth]{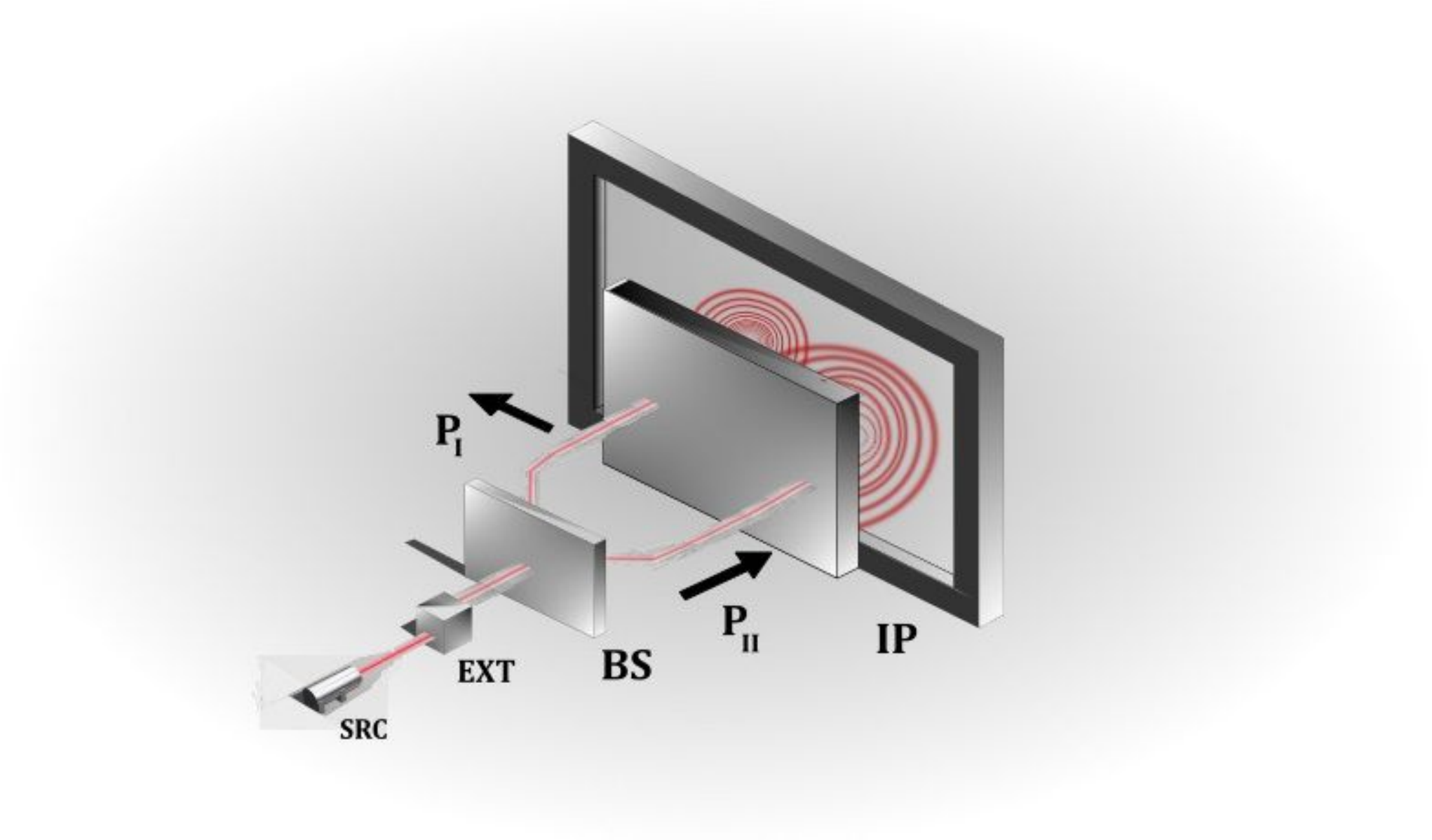}
\caption{
(color online) Schematic diagram of the interferometric apparatus. The beam from the neutron source \textbf{SRC} is conveyed to an external apparatus \textbf{EXT} consisting of a collimation device and a monochromator. The collimated and monochromatic beam is then conveyed to a beam splitter \textbf{BS} which splits the beam into two subbeams with different average spin polarizations $\pmb{P}_I$ and $\pmb{P}_{II}$. Finally the two subbeams interfere at the interference plane \textbf{IP}.
}
\label{IntDiagram}
\end{figure}

The setup is schematically pictured in figure (\ref{IntDiagram}). The fractional intensities of the subbeams with respect to the initial beam $I_0$, namely $\chi_I = \frac{I_{I}}{I_0}$ and $\chi_{II} =\frac{I_{II}}{I_0} $ should be as close as possible $\chi_I \simeq \chi_{II} $. They regulate the average distance between successive neutrons in the two sub--beams $d_{I}$ and $d_{II}$ as $I_{J} = \frac{\bar{v}_J}{d_J}$, with $\bar{v}_J$ the average neutron velocity in the direction of propagation for the $J$ sub--beam. Considering constant intensities $I_{J}$ and constant average neutron velocities $\bar{v}_J$, the average distances themselves $d_{J}$ are constant, resulting in a time--independent one particle Hamiltonian $H$, and if $\chi_I \simeq \chi_{II}$, $d_{I} \simeq d_{II}$.
The $d_J$ should be in any case large enough to neglect the effects of nuclear interactions $d_J > 10^{-12} m$. For simplicity, from now on, we assume that all the nearest neighbour distances within sub--beam $J$ are constant and equal to $d_J$.

The two polarized sub--beams then go through two optical paths of the same length to the interference plane \textbf{IP}, where the interference pattern is observed.
We choose $\pmb{P}_{I}$ to be orthogonal to $\pmb{\hat{y}}_I$  and $\pmb{P}_{II}$ parallel to $\pmb{\hat{y}}_{II}$. Starting from the total interaction Hamiltonian, exploiting the assumption that all the distances between two subsequent neutrons in sub--beam $J$ are equal to $d_J$, with simple algebraic steps, one arrives at the effective magnetic fields. It is easy to show that, within these assumptions, each neutron in sub--beam $J$ is subject to the same effective magnetic field $\mu \pmb{B_J} = \mu B_J \pmb{P_J}$ given by
\begin{eqnarray}\label{SecondSeries}\nonumber
\! \! \! \! \!   \mu B_I \! &\!= \! & \! \!  - \frac{2\mathcal{A}\zeta(3)}{d^3_I} \! + \! \frac{2\mathcal{A} \mathcal{B} }{d^3_I}  \mathrm{Li}_{3} (e^{-md_I} \!) \! + \! \frac{2\mathcal{A} \mathcal{B} m}{d^2_I}  \mathrm{Li}_{2} (e^{-md_I}\!)\\
\mu B_{II} \! &\!=\!&\! \frac{4\mathcal{A}\zeta(3)}{d^3_{II}}- \frac{4\mathcal{A} \mathcal{B} }{d^3_{II}}  \mathrm{Li}_{3} (e^{-md_{II}}) -
  \frac{4\mathcal{A} \mathcal{B} m}{d^2_{II}}\mathrm{Li}_{2} (e^{-md_{II}})  \nonumber \\
&& + \frac{2 \mathcal{A} \mathcal{B} m^2}{d_{II}} \log(1 - e^{-md_{II}})
\end{eqnarray}
where $\zeta(s)$ stands for the Riemann zeta function while $\mathrm{Li}_{s}(z) = \sum_{n=1}^{\infty} \frac{z^n}{n^s}$ is the Polylogarythm function \cite{Gradshteyn}.

Then the evolution of the single neutron state, in both sub--beams, is governed by the Schroedinger equation
\begin{equation}\label{Schrodinger}
  i \partial_t \psi_J = \left(- \frac{\nabla^2}{2M} + M \right) \psi_J - \pmb{\sigma} \cdot \left[\mu (\pmb{B_J} + \pmb{B^{0}_J} ) \right]\psi_J
\end{equation}
where $\psi_J$ is the product of a spatial wave-function (a plane wave $f(t) e^{i \pmb{k} \cdot \pmb{x}}$ if the beam is perfectly monochromatic) and a spin function. It is convenient to write the spinor for each sub--beam in the basis defined by the corresponding polarization, namely
$  \pmb{\sigma} \cdot \pmb{P_J} \ket{\uparrow_J} =  \ket{\uparrow_J}$ and $\pmb{\sigma} \cdot \pmb{P_J}\ket{\downarrow_J} = -\ket{\downarrow_J}.$
We assume that as soon as the neutron leaves \textbf{BS}, it is found in the up state for the corresponding sub--beam; at this instant we set $t=0$.

If $y$ denotes the coordinate along the propagation axis, with $y=0$ at the beginning of the optical path, thereby setting $t=0$, we have
$
  \psi_J (t) = f(t) e^{i k y} \ket{\uparrow_J}
$
where $f(t)$ is a function to be determined. Assuming $f(t) = e^{- i \omega_J t}$ and substituting in Eq.\eqref{Schrodinger} gives
$
  \omega_J = \frac{k^2}{2M} + M - \mu B_J - \mu B_0 \ .
$
Then the total phase accumulated in a time $t$ is equal to
$
  \phi_J (t) = \arg \left(\langle \psi_J(0)| \psi_J (t) \rangle\right) = - \left(\frac{k^2}{2M} + M - \mu (B_J + \mu B_0) \right)t.
$
The phase difference between the two beams at the interference plane is, after a time $t$,
$
  \Delta \phi (t) = \phi_{II} (t) - \phi_{I}(t) =\mu \left( B_{II} -  B_{I} \right)t
$
which can be computed immediately with the aid of Eq. \eqref{SecondSeries}. Assuming $d_{I}=d_{II} = d$, the result can be compactly expressed as $ \Delta \phi (t) = \left[G_{m} (d) + G_{a} (d)\right] t$, with $G_{m} (d)$ coming from the dipole--dipole interaction  and $G_{a} (d)$ from the axion--mediated interaction. These are $ G_{m} (d)  = \frac{6\mathcal{A}}{d^3} \zeta (3)$ and
\begin{eqnarray}\label{ThirdSeries}
\nonumber
G_{a} (d) &=&  -\frac{6\mathcal{A} \mathcal{B} }{d^3}  \mathrm{Li}_{3} (e^{-md}) - \frac{6\mathcal{A} \mathcal{B} m}{d^2}\mathrm{Li}_{2} (e^{-md})
\\
& +&\frac{2 \mathcal{A} \mathcal{B} m^2}{d} \log (1-e^{-md})  \,.\nonumber
\end{eqnarray}
where the $n \rightarrow \infty$ limit is understood.
To the phase difference $\Delta \phi (t)$ one must in principle add a possible phase shift $\Delta \phi_{BS}$ due to the operations conducted in the beam splitter and in the Stern--Gerlach like apparatus, which for our purposes should be as small as possible. To isolate the axion contribution in the phase difference, one can set the beam path (and then the evolution time $t$) in such a way that the first term, due to dipole-dipole interactions, is an integer multiple of $2 \pi$, since this phase difference is indistinguishable from a vanishing phase difference. These times, for each integer $k$, are clearly given by
$
  T_k = \frac{2 k \pi}{G_{m} (d)}= \frac{k \pi d^3}{3 \mathcal{A} \zeta(3)} \ ,
$
and the phase difference, evaluated at $T_k$, reads
\begin{eqnarray}\label{Reduced Phase Difference 2}\nonumber
 \nonumber \Delta \phi (T_k)  &=&   \bigg\{\frac{k \pi \mathcal{B}}{3 \zeta(3)} \Big[ 2 m^2 d^2 \log(1-e^{-md})
  \\
  &-& 6 m d \mathrm{Li}_2(e^{-m d}) - 6 \mathrm{Li}_3(e^{-md}) \Big]\bigg\}_{\mathrm{mod} \,2 \pi}.
\end{eqnarray}
Notice that this phase is non--zero only in presence of ALPs, since it vanishes for $\mathcal{B}=0$.
As it is evident from Eq.\eqref{Reduced Phase Difference 2}, the phase difference, evaluated at the $k$-th recurrence time $T_{k}$, is proportional to the parameter $\mathcal{B} \propto g_p^2$. In the range of ALP masses $m \in [10^{-6}-1]\mathrm{eV}$, and inter--neutron distances $d \in [10^{-11}-10^{6}]\mathrm{m}$, the phase depends only weakly on the parameters $m,d$, while keeping a relatively strong dependence on the coupling.
\begin{figure}[h]
\centering
\includegraphics[width=0.94\linewidth]{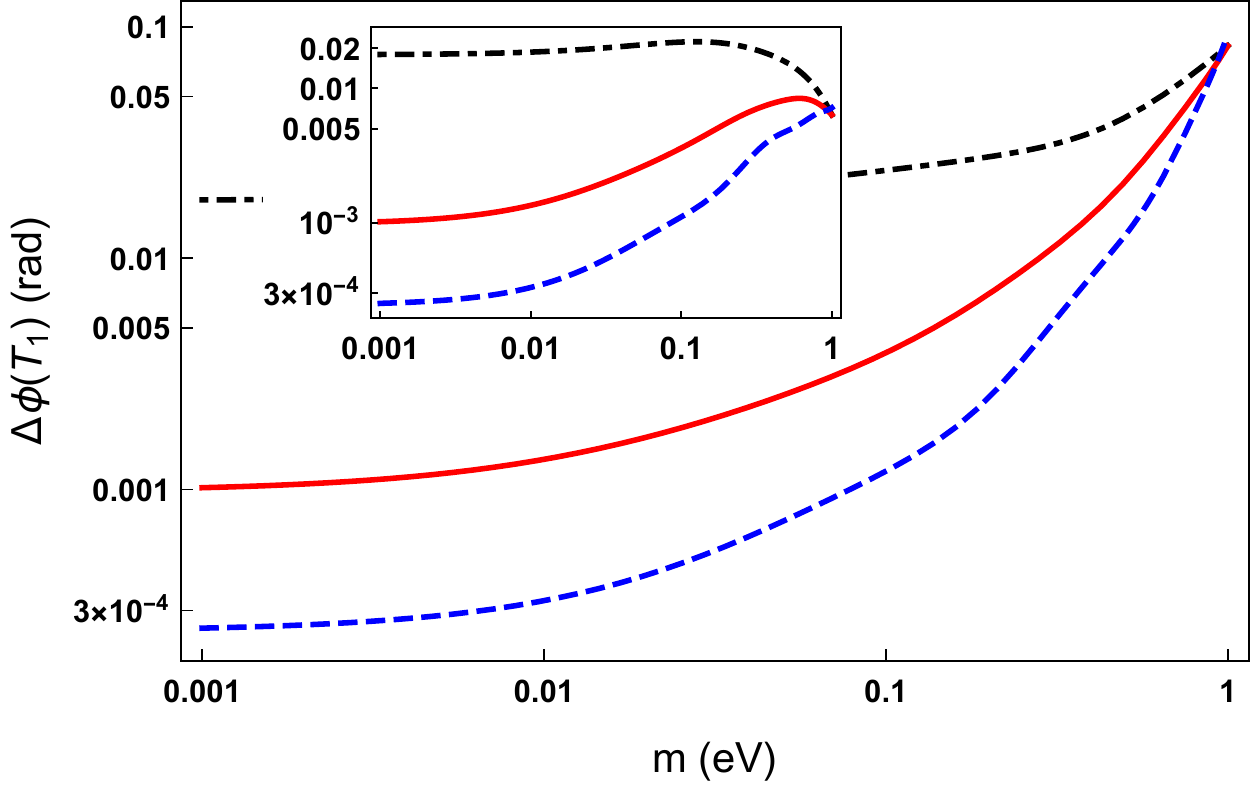}
\caption{
(color online) Logarithmic scale plot of the phase difference $|\Delta \phi (T_{1})|$ (Eq. \eqref{Reduced Phase Difference 2}) modulo $2 \pi$ for several values of the coupling constant in the ALP mass range $[10^{-3},1] \ \mathrm{eV}$ for an inter--neutron distance $d=10^{-8} \mathrm{m}$ ($d=10^{-6} \mathrm{m}$ in the inset). In particular:  (black dot--dashed line) $g_p = g_{CP}$, where $g_{CP}$ is the threshold from effective Casimir pressure measurements ~\cite{Bezerra2014} and sample values are $g_{CP} = 0.0327$ for $m = 10^{-3} \mathrm{eV}$, $g_{CP} = 0.0348$ for $m = 0.05 \mathrm{eV}$, $g_{CP} = 0.0674$ for $m = 1 \mathrm{eV}$; (red solid line) $g_p = g_{CF}$, where $g_{CF}$ is the threshold from measurements of the difference of Casimir forces ~\cite{Klimchitskaya2017} and sample values are $g_{CF}=0.007$ for $m = 10^{-3} \mathrm{eV}$, $g_{CF} = 0.012$ for $m = 0.05 \mathrm{eV}$, $g_{CF} = 0.066$ for $m = 1 \mathrm{eV}$; (blue dashed line) $g_p = g_{IE}$, where $g_{IE}$ is the threshold from isoelectronic experiments~\cite{Klimchitskaya2015}, and sample values are $g_{IE} = 0.0036$ for $m = 10^{-3} \mathrm{eV}$, $g_{IE}=0.006$ for $m = 0.05 \mathrm{eV}$, $g_{IE}=0.07$ for $m = 1 \mathrm{eV}$.
}
\label{Moste}
\end{figure}
In fig. (\ref{Moste}) we plot the phase difference, modulo $2 \pi$, evaluated at the minimum recurrence time $T_{1}$ for several values of the coupling constant $g_p$ in the mass range $[10^{-3},1] \ \mathrm{eV}$. For $m < 0.1 \mathrm{eV}$ we see that the phase difference is essentially the same for $d=10^{-8} \mathrm{m}$ and $d=10^{-6} \mathrm{m}$, showing that the dependence on the distance becomes relevant only when the product $md$ is quite high (right tail of the curves in the inset of (\ref{Moste})). This obviously traces back to the Yukawa damping factor $e^{-md}$ which accompanies the axion--mediated interaction, and can be clearly seen from Eq.\eqref{Reduced Phase Difference 2} when $ md\rightarrow 0$:
$
   \Delta \phi (T_k) \simeq \left[-\frac{2k \pi \mathcal{B}}{ \zeta(3)}\mathrm{Li}_3(1)\right]_{\mathrm{mod}\, 2 \pi}
$
, implying that when $md \ll 1$, the phase difference essentially depends only upon the coupling $g_p$ (via the parameter $\mathcal{B}$).

\begin{figure}
\centering
\includegraphics[width=\linewidth]{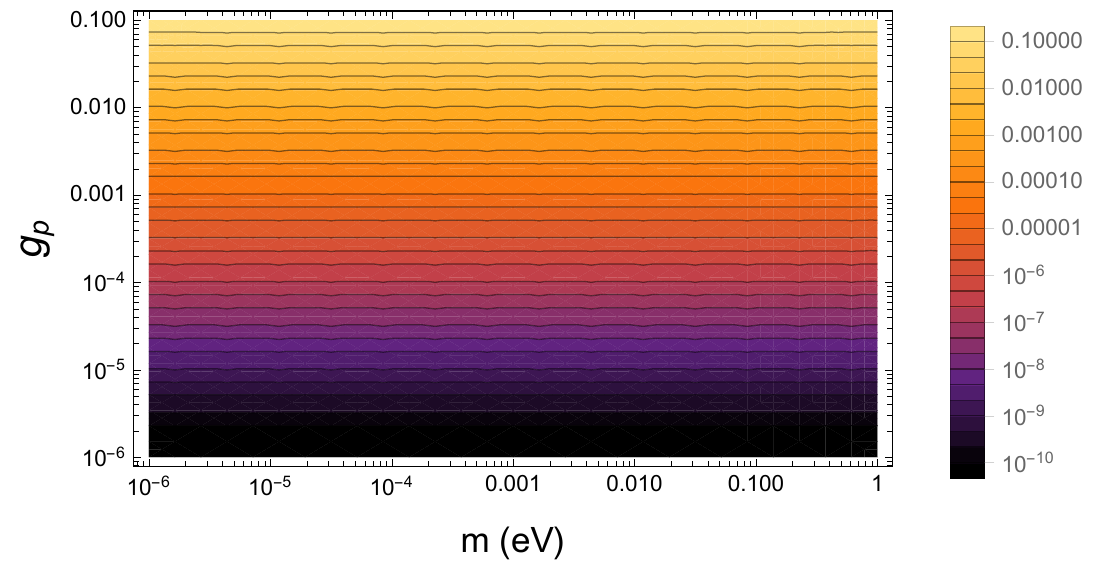}
\includegraphics[width=\linewidth]{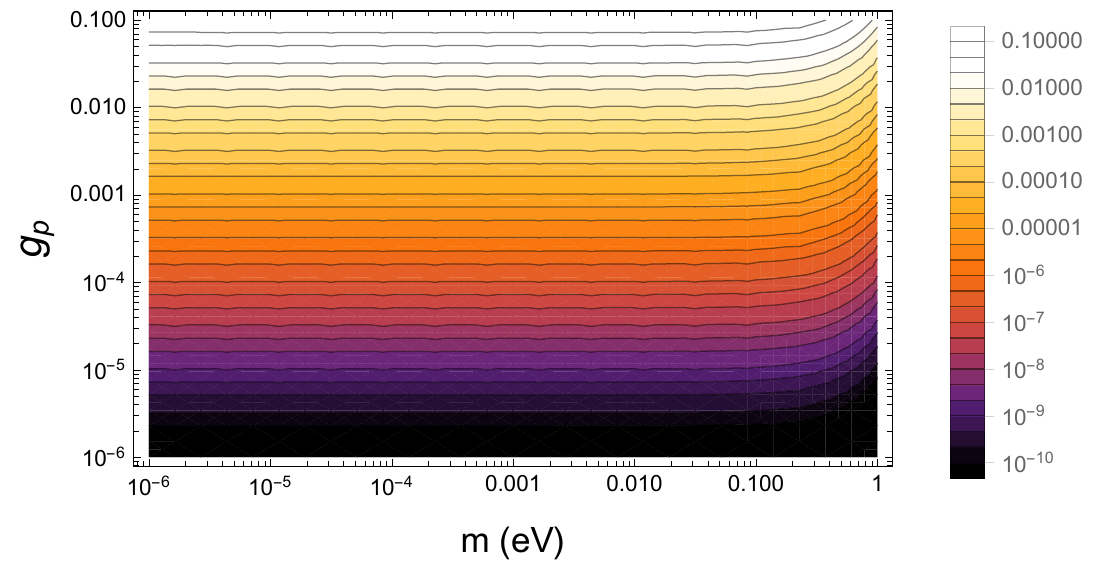}
\caption{
(color online) Contour plots of the phase difference $|\Delta \phi (T_{1})|$ in the mass--coupling plane for $d=10^{-9} m$ (upper panel) and $d=10^{-6} m$ (lower panel) in the range $(m,g_p) \in [10^{-6},1] \mathrm{eV} \times [10^{-6},10^{-1}]$. The Yukawa damping is visible in the lower panel at the right end of the plot, where curves corresponding to lower values of the phase difference are pushed up vertically.
}
\label{Contour}
\end{figure}

A similar conclusion can be drawn from figure (\ref{Contour}). Here the damping is evident for $d=10^{-6} \mathrm{m}$ (lower panel). Figure (\ref{Contour}) also shows clearly the $g_p^2$ dependence of the phase difference and an extremely weak dependence on the ALP mass in the ranges considered. It should be remarked that when the inter--neutron distance $d$ is small enough, one can evaluate the phase difference at the $k$-th recurrence time with a high $k$, while keeping the propagation time reasonable. This has the effect of bringing along a multiplicative $k$ factor in the phase difference (Eq. \eqref{Reduced Phase Difference 2}), which in principle may render it observable at lower values of the coupling constant. For instance, at $d=10^{-9} \mathrm{m}$ one can choose $k \simeq 4000$ while still keeping $T_{k} < 1 \mathrm{s}$. The observability of the phase difference is limited by the finite precision achievable in the experiment, and from the elements over which one has only a poor control. The first obvious limitation comes from the need for a recurrence time $T$  in order to isolate the axion contribution to the phase difference. This time cannot be arbitrarily long, because of the finite neutron lifetime $\tau_n \simeq 880 \mathrm{s}$ and especially because of a finite coherence length (and therefore a finite coherence time) which may be the result of unwanted interactions with the environment. Recurrence times of the order of the second can be obtained for $d \simeq 10^{-8} \ \mathrm{m}$. At the opposite end stands the issue of time resolution. A recurrence time too small would imply an extremely fast oscillation of the phase difference, rendering its analysis uneasy. This is not really a problem, unless very small inter--neutron distances $d < 10^{-11} \mathrm{m}$ are considered. The minimum recurrence time $T$ also has an obvious impact on the size of the interferometric apparatus needed. If $v$ denotes the average neutron velocity, the arm length must be at least $L=vT$. If $T \simeq \mathrm{s}$ and $L$ must be of the order of meters, the average neutron velocity cannot exceed a few $\mathrm{m/s}$. For this reason ultra cold neutrons are to be preferred, as they can have velocities as small as $5 \ \mathrm{m/s}$ \cite{Steyerl1969}. The second important limitation comes from the total neutron flux available, and therefore the available beam intensity $I$. Because of the several devices involved in the preparation of the beam (collimator, monochromator and so on) the flux loss may be significant. A relatively high intensity of the beam, on the other hand, is essential for a sufficiently small inter--neutron distance. In turn, a small $d$ guarantees a reasonable recurrence time. If the neutron velocity is of the order of $v \simeq 1 \ \mathrm{m/s}$, a distance $d \simeq 10^{-8} \mathrm{m}$ is obtained in correspondence with an intensity $I \simeq 10^{8} \mathrm{n/s}$. This is reasonable, given that pulsed neutron beams can reach intensities of the order $I \simeq 10^{10} n/s$ \cite{IAEA2014}.

In conclusion, we have  shown that the fermion-fermion interaction mediated by axions
 produces a non trivial contribution to  the total phase due to the time evolution  of neutrons. Then, we have considered a neutron interferometer with paths such that the interference  between the two beams depends only on the magnetic and axion--mediated interactions. We have shown that one can have a contribution to the  phase difference for the  dipole-dipole interaction which is an integer multiple of $2 \pi$
and then obtain a phase difference entirely due to the neutron-neutron interaction induced by axions.
By considering ultra cold neutrons, we derived values of phase differences which are in principle experimentally detectable for a wide range of ALP parameters.
Future experiments on neutron interferometry could be used to analyze the interaction induced by axions and to verify the existence of some types of ALPs.
In our treatment, for simplicity, we have modelled the collimated cold neutron beam as a 1--dimensional chain of neutrons which translates rigidly. We neglected the losses due to the
propagation, and considered the beam intensity as a constant. These approximations shall be improved upon in future works.

\section*{Acknowledgements}

A.C. and A.Q thank  partial financial support from MIUR and INFN. A.C. also thanks the COST Action CA1511 Cosmology and Astrophysics Network for Theoretical Advances and Training Actions (CANTATA).
SMG acknowledge support from the European Regional Development Fund for the Competitiveness and Cohesion Operational Programme (KK.01.1.1.06--RBI TWIN SIN) and from the Croatian Science Fund Project No. IP-2016--6--3347 and  IP-2019--4--3321.
SMG also acknowledge the QuantiXLie Center of Excellence, a project co--financed by the Croatian Government and European Union through the European Regional Development Fund--the Competitiveness and Cohesion Operational Programme (Grant KK.01.1.1.01.0004).


\begin{thebibliography}{99}



\bibitem{Ellis2009}
J. Ellis, \textit{Nucl. Phys. A} 827, 187-198 (2009).

\bibitem{Mixing1}
S. M. Bilenky and B. Pontecorvo, \textit{Phys. Lett. B} {\bf 61}, 248 (1976).

\bibitem{Mixing2}
S. M. Bilenky and B. Pontecorvo, \textit{Yad. Fiz.} {\bf 3}, 603 (1976).


\bibitem{Mixing3}
O. Nachtmann, {\it``Elementary Particle Physics:
Concepts and Phenomena''}, Springer, Berlin (1990).



\bibitem{Mixing4}
A. Capolupo, G. Lambiase, and A. Quaranta
\textit{Phys. Rev. D}, \textbf{101}, 095022 (2020).


\bibitem{dark1}
V. Rubin, W.K. Jr. Thonnard,  N.  Ford,
\textit{The Astrophysical Journal}, \textbf{238}: 471, (1980).


\bibitem{dark2}
N. Aghanim et al., Planck Collaboration,
\textit{Astron.Astrophys.} \textbf{641}, A6, (2020).

\bibitem{Quinn1977}
R. D. Peccei and H. Quinn, \textit{Phys. Rev. Lett.} {\bf 38}, 1440 (1977).

\bibitem{Peccei1977}
R. D. Peccei and H. Quinn, \textit{Phys. Rev. D} \textbf{16} 1791 (1977).

\bibitem{Weinberg1978}
S. Weinberg, \textit{Phys. Rev. Lett.} \textbf{40}, 223 (1978).

\bibitem{Wilczek1978}
F. Wilczek, \textit{Phys. Rev. Lett.} \textbf{40}, 279 (1978).

\bibitem{Raffelt2007}
G. G. Raffelt, \textit{J. Phys. A} \textbf{40}, 6607 (2007).

\bibitem{Marsch2016}
D. J. E. Marsch \textit{Phys. Rep.}, \textbf{643}, 1 (2016).


\bibitem{DFSZ1}
M. Dine, W. Fischler, and M. Srednicki, Phys. Lett. B \textbf{104}, 199 (1981).

\bibitem{DFSZ2}
A. Zhitnitsky, Sov. J. Nucl. Phys. \textbf{31}, 260 (1980).

\bibitem{PDG2020}
Particle Data Group P. A. Zyla et al., \textit{Progress of Theoretical and Experimental Physics}, vol. \textbf{2020}, Issue 8, August 2020, 083C01 (2020).

\bibitem{KSVZ1}
J. E. Kim, Phys. Rev. Lett. \textbf{43}, 103 (1979).

\bibitem{KSVZ2} M. A. Shifman, A. Vainshtein, and V. I. Zakharov,
Nucl. Phys. B 1\textbf{66}, 493 (1980).

\bibitem{Kim2016}
J. E. Kim and D. J. E. Marsch, \textit{Phys. Rev. D} \textbf{93}, 025027 (2016).

\bibitem{Demartino2017}
I. De Martino, T. Broadhurst, S.-H. Henry-Tye, T. Chiueh, H.-Y. Schive and R. Lazkoz, \textit{Phys. Rev. Lett.} \textbf{119}, 221103 (2017).

\bibitem{Rubakov1997}
V. A. Rubakov, \textit{JETP Lett.} \textbf{65}, 621-624 (1997).



\bibitem{PVLAS}
E. Zavattini et al. (PVLAS Collaboration), \textit{Phys. Rev. D} \textbf{77}, 032006 (2008).

\bibitem{OSQAR}
P. Pugnat et al. (OSQAR Collaboration), \textit{Phys. Rev. D} \textbf{78}, 092003 (2008).
\bibitem{OSQAR2}
R. Ballou et al. (OSQAR Collaboration), \textit{Phys. Rev. D} \textbf{92}, 092002 (2015).

\bibitem{ALPS}
K. Ehret et al. (ALPS Collaboration), \textit{Phys. Lett. B} \textbf{689}, issues 4-5, pages 149-155 (2010).

\bibitem{CAST}
S. Aune et al. (CAST Collaboration), \textit{Phys. Rev. Lett.} \textbf{107}, 261302 (2011).

\bibitem{ADMX}
N. Du et al. (ADMX Collaboration), \textit{Phys. Rev. Lett.} \textbf{120}, 151301 (2018).

\bibitem{QUAX}
R. Barbieri et al., \textit{Phys. Dark Univ.} \textbf{15}, 135-141 (2017).

\bibitem{Capolupo2015}
A. Capolupo, G. Lambiase, G. Vitiello, \textit{ Adv. In High En. Phys.} 826051 (2015).

\bibitem{Capolupo2019}
A. Capolupo, I. De Martino, G. Lambiase, A. Stabile, \textit{Phys. Lett. B} \textbf{790}, 427-435 (2019).




\bibitem{Moody1984}
J. E. Moody, F. Wilczek, \textit{Phys. Rev. D}, \textbf{30}, 130 (1984).
\bibitem{Daido2017}
R. Daido, F. Takahashi,  \textit{Phys. Lett. B}, \textbf{772}, 127 (2017).


\bibitem{Bezerra2014}
V. B. Bezerra, G. L. Klimchitskaya, V. M. Mostepanenko and C. Romero, \textit{Eur. Phys. J. C} \textbf{74}, 2859 (2014).

\bibitem{Klimchitskaya2017}
G. L. Klimchitskaya and V. M. Mostepanenko, \textit{Phys. Rev. D} \textbf{95} 123013 (2017).

\bibitem{Klimchitskaya2015}
G. L. Klimchitskaya and V. M. Mostepanenko, \textit{Eur. Phys. J. C} \textbf{75}, 164 (2015).






\bibitem{Capolupo2020}
A. Capolupo, S. M. Giampaolo, G. Lambiase, A. Quaranta, \textit{Phys. Lett. B} \textbf{804}, 135407 (2020).


\bibitem{Rauch2015}
H. Rauch, S. A. Werner, \textit{Neutron Interferometry--Lessons in Experimental Quantum Mechanics, Wave--Particle Duality, and Entanglement}, 2nd Edition, Oxford University Press, Oxford (2015).

\bibitem{Werner1979}
S. A. Werner, J.--L. Staudenmann, R. Colella, \textit{Phys. Rev. Lett.}, \textbf{42}, 1103 (1979).


\bibitem{Allman1997}
B. E. Allman, H. Kaiser, S. A. Werner, A. G. Wagh, V. C. Rakhecha, J. Summhammer, \textit{Phys. Rev. A}, \textbf{56}, 4420 (1997).


\bibitem{Wagh1990}
A. G. Wagh, V. C. Rakhecha, \textit{Phys. Lett. A}, \textbf{148}, Issues 1--2, pp. 17-19 (1990).


\bibitem{Wagh1998}
A. G. Wagh, V. C. Rakhecha, P. Fischer, A. Ioffe, \textit{Phys. Rev. Lett.}, \textbf{81}, 1992 (1998).


\bibitem{Ott2015}
F. Ott, S. Kozhevnikov, A. Thiaville, J. Torrejon, M. Vazquez, \emph{Nucl. Inst. Meth. in Phys. Res. A}, \textbf{788} (2015), pp. 29-34.

\bibitem{Gradshteyn}
I. S. Gradshteyn, I. M. Ryzhik, \textit{Table of Integrals, Series and Products}, 7th edition, Academic Press (2007), pp. 110-111.





\bibitem{Steyerl1969}
A. Steyerl, \textit{Phys. Lett. B}, \textbf{29}, Issue 1, pp. 33-35 (1969).

\bibitem{IAEA2014}
INTERNATIONAL ATOMIC ENERGY AGENCY, \textit{Compendium of Neutron Beam Facilities for High Precision Nuclear Data Measurements}, IAEA-TECDOC-1743, IAEA, Vienna (2014).


\end{thebibliography}
\end{document}